\documentclass[10pt,aps,prl,twocolumn,floatfix, tightenlines,showpacs,showkeys,preprintnumbers,footinbib, longbibliography]{revtex4-1}
\pdfoutput=1
\usepackage[colorlinks=true,citecolor=blue,linkcolor=blue,breaklinks=true]{hyperref}
\usepackage{amsmath,amssymb,mathtools,tabu}
\usepackage{epsfig}  
\usepackage{graphicx}   
\usepackage{commath}
\usepackage{slashed}             
\usepackage{url}
\usepackage{color}
\usepackage{multirow}
\usepackage{placeins}
\usepackage[dvipsnames]{xcolor}

\allowdisplaybreaks

\bibpunct{[}{]}{,}{n}{}{,}

\begin{document}

\title{Selection Rule for Enhanced Dark Matter Annihilation}

\author{Anirban~Das}
\email{anirbandas@theory.tifr.res.in }              
\author{Basudeb~Dasgupta}
\email{bdasgupta@theory.tifr.res.in}
\affiliation{Tata Institute of Fundamental Research,
             Homi Bhabha Road, Mumbai, 400005, India.}

\preprint{TIFR/TH/16-43}
\pacs{95.35.+d} 
\date{June 29, 2017}

\begin{abstract}
We point out a selection rule for enhancement (suppression) of odd (even) partial waves of dark matter coannihilation or annihilation using Sommerfeld effect. Using this, the usually velocity-suppressed \mbox{$p$-wave} annihilation can dominate the annihilation signals in the present Universe. The selection mechanism is a manifestation of the exchange symmetry of identical incoming particles, and generic for multi-state DM with off-diagonal long-range interactions.  As a consequence, the relic and late-time annihilation rates are parametrically different and a distinctive phenomenology, with large but strongly velocity-dependent annihilation rates, is~predicted.

\end{abstract}

\maketitle

{\bf \emph{1.\,Introduction.--}} 
If DM is a thermal relic of the early Universe, then its cosmological abundance provides a measure of its ``annihilation rate" $\langle\sigma v\rangle$\,\cite{Steigman:1984ac, Jungman:1995df, Bertone:2004pz}. This annihilation rate has contributions from various partial waves of the scattering amplitude, each with its characteristic dependence on the relative velocity $v$ of the colliding particles,
\begin{align}
 \langle \sigma v \rangle = \underbrace{\,a\,}_{{\,}_{\langle \sigma v \rangle_s}} + \underbrace{\,b v^2\,}_{{\,}_{\langle \sigma v \rangle_p}} +\,\ldots\,.
\end{align}
The first term on the right, $\langle \sigma v \rangle_s$, represents the velocity-independent \mbox{$s$-wave} contribution  and the second term, $\langle \sigma v \rangle_p$, which scales as $v^2$, has the \mbox{$p$-wave} contribution. Omitted terms appear with higher powers of $v^2$, and for nonrelativistic DM, the contribution of these higher partial waves are small. In the simplest models, the \mbox{$s$-wave} contribution dominates and the annihilation rate is $\langle\sigma v\rangle^{\rm relic}\simeq2.2\times10^{-26}\,{\rm cm^3s^{-1}}$\,\cite{Steigman:2012nb}, to produce the observed DM abundance, practically independent of $v$.

Detection of a non-\mbox{$s$-wave} DM annihilation rate, e.g., $\langle\sigma v\rangle\propto v^2$, would reveal a crucial clue to the nature of DM. However, it is believed to be highly challenging. To the best of our knowledge, annihilations of very dense or very fast DM are the only avenues that have interesting sensitivity to \mbox{$p$-wave} annihilations\,\cite{Diamanti:2013bia, Essig:2013goa, Drlica-Wagner:2014yca, Shelton:2015aqa}. Unfortunately, even these become inefficient for heavy DM.

Sommerfeld effect induces further nontrivial velocity-dependence of the annihilation rate\,\,\cite{ANDP:ANDP19314030302,Hisano:2003ec}. 
Long-range interactions of DM distort the wave-functions of incoming particles and change the annihilation rate, $\langle \sigma v \rangle\to S\langle \sigma v \rangle$, by the velocity-dependent Sommerfeld factor $S$. This effect has been studied extensively in recent years\,\mbox{\cite{Ibe:2008ye, Iengo:2009ni, Cassel:2009wt, Slatyer:2009vg,  McDonald:2012nc, Liu:2013vha, Blum:2016nrz, An:2016kie, Petraki:2016cnz, Asadi:2016ybp}}, after it was initially invoked\,\cite{Cirelli:2007xd, ArkaniHamed:2008qn,Pospelov:2008jd} to explain the cosmic-ray positron excesses\,\cite{Adriani:2008zr, Atic:2008} using a large DM annihilation rate. As this enhancement occurs for small $v$, again models with dominantly \mbox{$s$-wave} annihilations are popular. Therein, the enhancement is always larger for smaller $v$ and as a result a large annihilation rate is predicted around recombination, which leaves an imprint on the Cosmic Microwave Background (CMB)\,\cite{Slatyer:2009yq, Galli:2011rz}.

In this \emph{Letter}, we point out a selection mechanism that allows enhanced \mbox{\mbox{$p$-wave}} DM annihilation, with no enhancement but rather a possible suppression of the \mbox{$s$-wave} rate.  
Models employing the mechanism are testable and predict a distinctive, large but strongly velocity-dependent, annihilation rate: highest at intermediate velocities, e.g., $v\simeq10^{-3}$--$10^{-4}$ in galaxies, while being lower at both larger and smaller velocities, e.g., in galaxy clusters and at recombination, respectively. In the following, we explain this mechanism, provide a concrete model, discuss the main signatures and constraints, and finally conclude.

{\bf\emph{2.\,Mechanism.--}} The basic idea is that, for coannihilations or annihilations of multi-level DM, the effective one-level interaction potential can be attractive or repulsive, depending on the angular momentum of the incoming state, and leads to enhancement  or suppression, respectively. We now explain this \emph{selection mechanism} in more detail, for coannihilations or annihilations of two DM fermions $A$ and $B$.

Let $\Psi_i$ be the wave-function of an incoming two-body state, i.e., $|A B\rangle$ or $|B A\rangle$ for co-annihilation and $|A A\rangle$ or $|B B\rangle$ for annihilation, with the state labeled by $i\in\{1,2\}$ in each case.
Its long-distance distortion is governed by the two-level Schr\"odinger equation\,
\begin{align}
-\frac{1}{2\mu_i}\frac{{\rm d\Psi_i}^2}{{\rm d}r^2}+\frac{\ell(\ell+1)}{2\mu_i r^2}\Psi_i+V_{ij}(r)\Psi_j=\frac{k^2_i}{2\mu_i}\Psi_i\,,
\label{eq:schreqn}
\end{align}
where $k_i = \mu_i v$, with $\mu_i$ being the reduced mass of the $i$-th two-body state, $\ell$ is the angular momentum, and
\begin{align}
\label{pot1}
 V = \begin{pmatrix}
   V_{11} & V_{12}\\
  V_{21} &  V_{22}
 \end{pmatrix}\,,
\end{align}
the potential energy matrix dependent on interactions.  The Sommerfeld factor for coannihilation or annihilation channel $i$ and partial wave $\ell$ is given by\,\cite{Slatyer:2009vg}
\begin{align}
 S_\ell^{(i)} = \left(\frac{(2\ell-1)!!}{k_i^\ell}\right)^2 \frac{\left(T^\dagger \Gamma_{\ell} T\right)_{ii}}{(\Gamma_\ell)_{ii}}\,\quad{\rm (no\ sum)}\,.
\label{eq:Sfactor}
\end{align}
The matrix {$T_{ij}=\Psi_i^{*}\Psi_j e^{-ik_i r}|_{r \to \infty}$} consists of the amplitudes of asymptotic wavelike solutions to eq.\,(\ref{eq:schreqn}). The $\Gamma_\ell$-matrix contains the $a,\,b,\ldots$ coefficients of coannihilation or annihilation rates, calculable in the framework of non-relativistic effective theory of the DM model\,\cite{Beneke:2012tg, Hellmann:2013jxa, Beneke:2014gja}.

{\emph{Equivalent One-level Problem.--\,}} For \emph{co-annihilation}, physically there is no distinction between the two states, $|A B\rangle$ and $|B A\rangle$, as one is obtained from the other by a mere \emph{exchange} of particles. Therefore, one expects these states to be identical up to an overall phase,
\begin{align}
 |BA\rangle = (-1)^{\ell + s}|AB\rangle\,,
\label{eq:exchange}
\end{align}
where $\ell, s$ are the angular momentum and spin of the two-body state\,\cite{Beneke:2012tg}. A factor of $(-1)^\ell$ comes from the change in relative momentum, $(-1)^{s+1}$ from exchange of spins, and a $(-1)$ from the Wick exchange of fermion fields. Clearly, the potentials must satisfy $V_{11}=V_{22}$ and $V_{12}=V_{21}$. Plugging eq.\,(\ref{eq:exchange}) in eq.\,(\ref{eq:schreqn}), reduces eq.\,(\ref{eq:schreqn}) to its one-level-equivalent with the effective potential
\begin{align}
V_{\rm eff}=V_{11}+(-1)^{\ell+s}V_{12}+\frac{\ell(\ell+1)}{2\mu_i r^2}\,.
\label{eq:Veff}
\end{align}

The effective potential $V_{\rm eff}$ leads to selective Sommerfeld enhancement of odd or even partial waves. Consider, for example, the potentials $V_{ij}$ are attractive and that the incoming state has $s=1$. For even-integer values of $\ell$, e.g., \mbox{$s$-wave}, the effective interaction $V_{11}-V_{12}$ may vanish if $V_{12}\simeq V_{11}$ or become \emph{repulsive} if $|V_{12}|\gtrsim|V_{11}|$. Thus one expects no enhancement or perhaps even a suppression of the \mbox{$s$-wave} rate. On the other hand, for odd-integer values of $\ell$, e.g., \mbox{$p$-wave}, the potential is $V_{\rm eff}=V_{11}+V_{12}+\ell(\ell+1)/(2\mu_i r^2)$, which can be \emph{attractive} if $V_{11}+V_{12}$ falls off slower than $1/r^2$ in the relevant range. A minimum in the potential then develops at finite nonzero $r$, where higher $\ell$ wave-functions peak, and leads to an enhancement. As a result, one has $S_{\ell={\rm even}}\lesssim1$ and $S_{\ell={\rm odd}}\gg1$. If $|V_{12}|\ll|V_{11}|$, this mechanism is not as effective, $V_{\rm eff}$ being dominated by the diagonal potential that does not switch its sign. The general lesson here is that a strong off-diagonal long-range interaction of multi-level DM can enforce a spin/angular-momentum-dependent selection rule on Sommerfeld enhancement.

\begin{figure}[t]
\includegraphics[width=0.9\columnwidth]{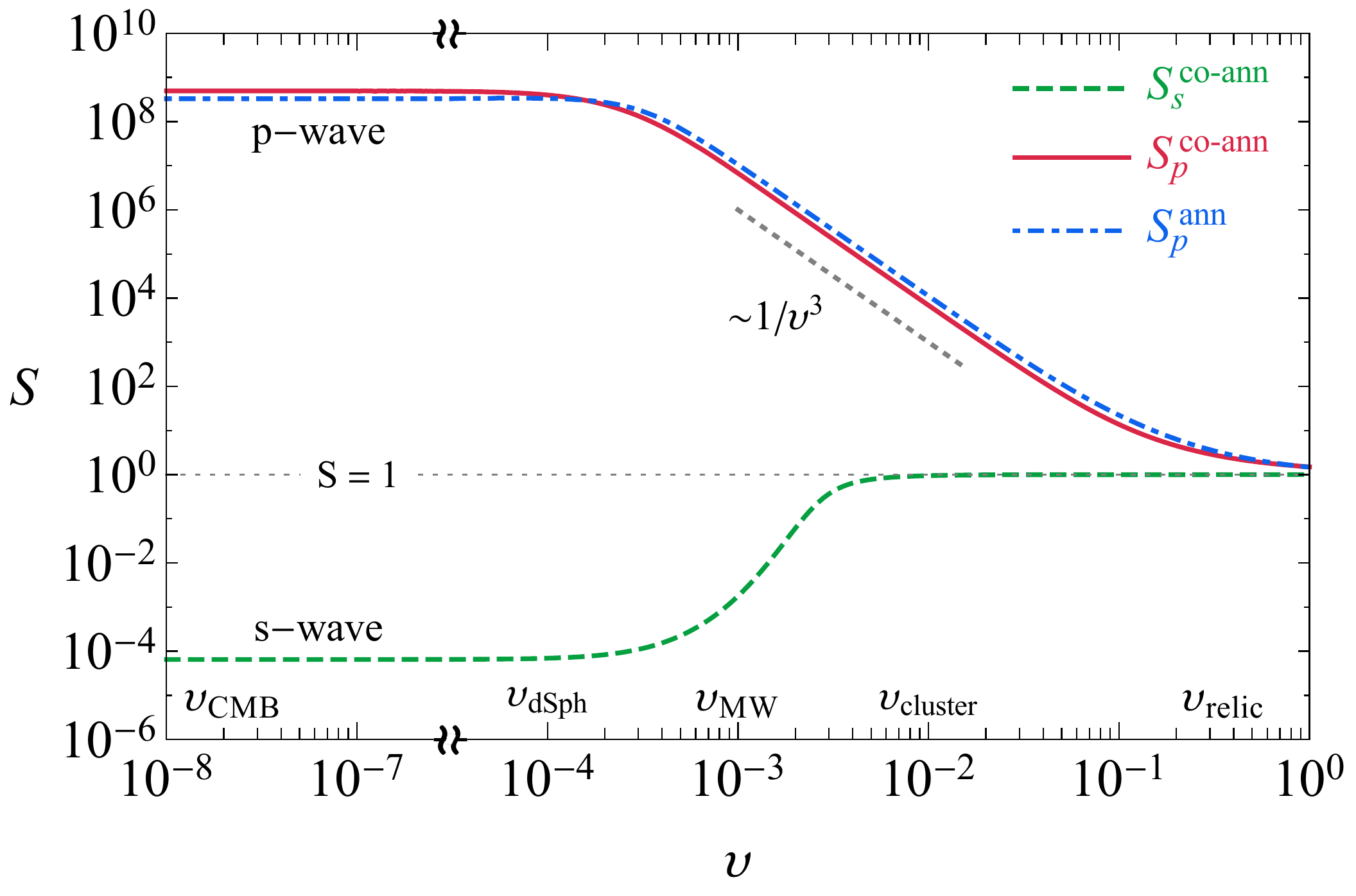}
\caption{Sommerfeld factors for \mbox{$s$-wave} and \mbox{$p$-wave} coannihilation or annihilation processes at velocities $v$. At smaller velocities the \mbox{$p$-wave} coannihilation or annihilation processes are strongly enhanced but \mbox{$s$-wave} co-annihilation is suppressed. The $\sim$$1/v^{3}$-rise of $S_p$ at intermediate velocities predicts that $\langle\sigma v\rangle\simeq S_{p}\langle\sigma v\rangle_{p}$ peaks for $v$ at the edge of the saturation plateau at low $v$. Typical DM velocities in different sources/epochs are annotated. See text for details of the model.}
\label{fig:1}
\end{figure}

Figure\,\ref{fig:1} shows a typical manifestation of this selection mechanism. At high velocities, $v\simeq1$, the Sommerfeld factors are close to 1, not appreciably affecting relic annihilation. At smaller velocities, \mbox{$s$-wave} rates are suppressed but the \mbox{$p$-wave} rates are enhanced, i.e., $S_s\lesssim1$ and $S_p\gg1$.  Specifically for $v \lesssim 10^{-3}{-}10^{-4}$, the \mbox{$p$-wave} Sommerfeld factors $S_p^{\rm co/ann}$ saturate to large constant values, rising roughly as $\sim 1/v^{3}$ in the intermediate region. This stronger velocity dependence for intermediate $v$ can overcome the $v^2$ suppression in $\langle \sigma v\rangle_p$ and produces a unique phenomenology. 

\break

For \emph{annihilation}, where the states $|A A\rangle$ and $|B B\rangle$ are not obviously related, the one-level-equivalent does not exist. Yet, as we see in Fig.\,\ref{fig:1}, the \mbox{$p$-wave} {annihilation} also shows a large enhancement. We will show that this is a consequence of an approximate $|A A\rangle\leftrightarrow|B B\rangle$ exchange symmetry, which when exact makes $|A A\rangle$ and $|B B\rangle$ identical to each other. Then, the preceding argument applies for annihilation as well, with small corrections proportional to the breaking of this symmetry.

{\bf\emph{3.\,Model.--}} The above selection mechanism or its variants will crop up in many existing DM models. For example, multiple DM fermions universally coupled to a boson in the Standard Model (SM) naturally exhibit the selection mechanism. Here, we discuss a simple model that presents an interesting version of the selection mechanism, where the late-time signal can be due to a \emph{purely} \mbox{$p$-wave} process.

Consider a Dirac fermion $\chi$ and a complex scalar $\phi$, with charges $+1$ and $-2$, respectively, under an explicitly broken global dark U(1) symmetry~\cite{Weinberg:2013kea, Garcia-Cely:2013nin, Chu:2014lja},
\begin{align}
 \mathcal{L} &\supset  \partial^\mu \phi^\dagger \partial_\mu \phi + \mu^{2} \abs{\phi}^2 - \lambda \abs{\phi}^4 +\mathcal{L}_{U(1)-{\rm breaking}}\nonumber \\
   &\quad + i \overline{\chi} \slashed{\partial}\chi - M \overline{\chi}\chi - \left(\frac{f}{\sqrt{2}} \phi \overline\chi\chi^c + h.c. \right)\,.
\label{eq:Lag}
\end{align}
$\phi$ develops a vacuum expectation value $v_\phi$, to give {$\phi=(v_\phi+\rho+i\eta)/\sqrt{2}$} and splits $\chi$ into two pseudo-Dirac DM particles $\chi_1=(\chi-\chi^c)/(\sqrt{2}i)$ and $\chi_2=(\chi+\chi^c)/\sqrt{2}$ with masses $M\mp\Delta/2$. Taking $\mathcal{L}_{U(1)-{\rm breaking}}=-\frac{1}{2}m_\eta^2\eta^2$ keeps the residual $\mathcal{Z}_2$-symmetry, which stabilizes the lighter $\chi_1$ of mass $m_\chi$ and makes it a good DM candidate while $\eta$ becomes a pseudo-Nambu-Goldstone boson of mass $m_\eta$. 

\pagebreak
The fermion interactions are $-\frac{f}{2}\rho\,(\overline{\chi}_1\chi_1-\overline{\chi}_2\chi_2)-\frac{f}{2}\eta\,(\overline{\chi}_1\chi_2+\overline{\chi}_2\chi_1)$, i.e., $\eta$ only mediates between different fermions, while $\rho$ mediates between alike fermions. The interaction potentials are then given by $V_{11} = -\alpha e^{-m_\rho r}/r, V_{12} = V_{21} = -\alpha e^{-m_\eta r}/r$, $V_{22} = -\alpha e^{-m_\rho r}/r$ for co-annihilation and $V_{22} = -\alpha e^{-m_\rho r}/r + 2\Delta$ for annihilation, with the dark fine-structure constant $\alpha \equiv f^2/(4\pi)$. A chiral fermion $\chi_L$ instead of $\chi$ in eq.\,(\ref{eq:Lag})~\mbox{\cite{Garcia-Cely:2013wda}}, would have led to a spin-dependent singular potential mediated by $\eta$ and the Sommerfeld effects would be very different~\cite{Bedaque:2009ri, Bellazzini:2013foa}. This problem does not arise here. We will be interested in the parameter space where \mbox{$m_\eta, m_\rho, \Delta \ll m_\chi$}. 

{\bf\emph{4.\,Methods \& Results.--}} The co-annihilation process has both \mbox{$s$-wave} and \mbox{$p$-wave} amplitudes, while for annihilation the \mbox{$s$-wave} process is forbidden by having identical Majorana fermions in the initial state\,\cite{Weinberg:2013kea, Garcia-Cely:2013nin, Garcia-Cely:2013wda, Chu:2014lja}. To compute the Sommerfeld factors for $\langle \sigma v\rangle^{\rm co{\textrm-}ann}_{s,p}$ and $\langle \sigma v\rangle^{\rm ann}_{p}$ , using eq.\,(\ref{eq:Sfactor}), following refs.\,\cite{Beneke:2012tg, Hellmann:2013jxa, Beneke:2014gja} we first computed $\Gamma_\ell$:
\begin{align}
\Gamma^{\rm co\textrm{-}ann}_{s} = \frac{\pi \alpha^2}{3 m_\chi^2} \begin{pmatrix}
                                                                 +1 & -1\\
                                                                 -1 & +1
                                                                \end{pmatrix},\\
\Gamma^{\rm co\textrm{-}ann}_{p} = \frac{\pi \alpha^2 m^4_\rho}{4 m_\chi^4 \Delta^2} \begin{pmatrix}
                                                                 +1 & +1\\
                                                                 +1 & +1
                                                                \end{pmatrix},\\
\Gamma^{\rm ann}_{p} = \frac{6\pi \alpha^2}{m_\chi^2} \begin{pmatrix}
                                                                 +1 & +1\\
                                                                 +1 & +1
                                                                \end{pmatrix}\,,
\end{align}
to leading order in $m_\rho^2/(m_\chi\Delta)\ll1$, i.e., when $\chi_{1,2}$ are not overly degenerate\,\cite{SuppMat}.
We then computed $T$ by numerically solving the corresponding Schr\"odinger equations [eq.\,(\ref{eq:schreqn})] using two methods: (i) directly solving the two-level equations\,\cite{Slatyer:2009vg}, and (ii) using the variable phase method with an ansatz for the wave-functions in terms of Bessel functions\,\cite{Ershov:2011zz,Beneke:2014gja}. The second method is especially useful in cases with exponentially growing solutions. For co-annihilation, where the two-level system can be exactly mapped into a one-level system, we computed the factor using the one-level equation as well\,\cite{Bellazzini:2013foa, Archidiacono:2014nda}. All methods gave identical results. 

Figure\,\ref{fig:1} shows the dependence of $S$ on velocity, for $m_\rho=10^{-3}\,m_\chi$, $m_\eta=0.9\,m_\rho$, $\Delta=10^{-3}m_\chi$, and $\alpha=0.1$ as representative values. The main feature, i.e., the enhancement of \mbox{$p$-wave} rates and suppression of the \mbox{$s$-wave} rate, is understood in terms of the effective potential.

{\emph{One-level Interpretation.--}} The \emph{co-annihilating} $|\chi_1\chi_2\rangle$ or $|\chi_2\chi_1\rangle$ states have total spin $s = 1$. The equivalent one-level problem then has the effective potential,
\begin{align}
 V_{\rm eff} =  -\frac{\alpha\,e^{-m_\rho r}}{r}+ (-1)^\ell\,\frac{\alpha\,e^{-m_\eta r}}{r}+ \frac{\ell(\ell+1)}{2\mu_i r^2}\,.
\label{eq:veff2}
\end{align}
The $\eta$ gives a Yukawa potential that being stronger than $1/r^2$ at $r\,{\sim}\,m_\eta^{-1}$ obviously satisfies the condition for selective enhancement. As expected from the general selection rule, the $\ell$-dependent sign of the second Yukawa potential leads to $S_p^{\rm co{\textrm-}ann}\gg1$ and $S_s^{\rm co{\textrm-}ann}\lesssim1$. 


\begin{figure}[t]
\includegraphics[width=0.8\columnwidth]{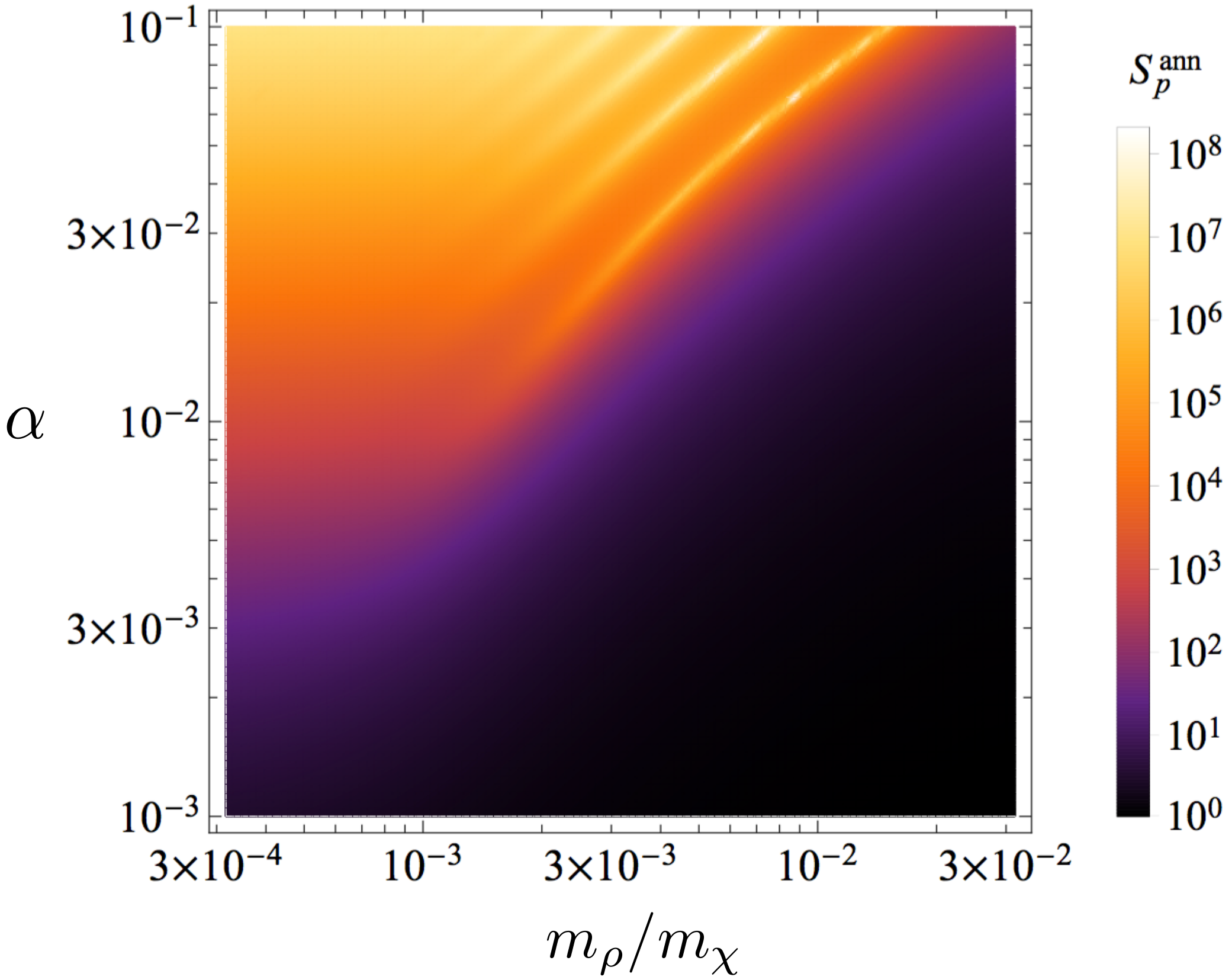}
\caption{Sommerfeld enhancement of \mbox{$p$-wave} annihilation, for DM of mass $m_\chi$ interacting via a mediator of mass $m_\rho$ with dark fine-structure constant $\alpha$. The Sommerfeld factor is large for $\alpha\gtrsim{\rm few}\times10^{-3}$ and $m_\rho/m_\chi\lesssim{\rm few}\times10^{-2}$, using $v=10^{-3}$ and $\Delta/m_\chi=10^{-3}$ as representative values.}
\label{fig:2}
\end{figure}
The \emph{annihilating} $|\chi_1\chi_1\rangle$ or $|\chi_2\chi_2\rangle$ states have spin $s = 1$, with $\ell+s$ being an even-integer due to antisymmetry. Thus, $\ell$ is odd.  This two-level problem does not reduce to a one-level problem directly. However, in the limit $\Delta\ll m_\chi$ one has $V_{11}=V_{22}$ and $|\chi_1\chi_1\rangle\leftrightarrow|\chi_2\chi_2\rangle$, and 
only the linear combination $|\chi_1\chi_1\rangle-(-1)^\ell|\chi_2\chi_2\rangle$ is physically relevant. For this linear combination, $V_{\rm eff}$ is the same as in eq.\,(\ref{eq:veff2}) and one gets $S^{\rm ann}_p\gg1$, i.e., p-wave \emph{annihilation} is also strongly Sommerfeld-enhanced; the physical origin of the enhancement being the \emph{approximate} exchange symmetry at $\Delta\to0$. 

As $\chi_2$ may decay to $\chi_1$, we focus on \emph{annihilations} at late time. Figure\,\ref{fig:2} illustrates the dependence of $S_p^{\rm ann}$ on the strength and range of the interaction. Here, $S_p^{\rm ann}\gg1$, when $\alpha\gtrsim10^{-3}$ and $m_\rho/m_\chi\lesssim{\rm few}\times10^{-2}$. At small  $\alpha$ there is no significant enhancement, whereas larger enhancements are possible when the momentum in the first Bohr orbit, $\sim$$\alpha m_\chi$, becomes larger than the relative momentum of incoming particles $\sim$$m_\chi v$\,\cite{Hoyer:2014gna}. For large $m_\rho$, the Yukawa potential is negligible and $S_p^{\rm ann}\to1$, whereas in the small $m_\rho$ limit, the potential and the solution become independent of $m_\rho$. For intermediate values of $m_\rho\simeq {6\alpha m_\chi}/({\pi^2 (n+2)^2})$, with $n = 0,1,2,\ldots$\,\cite{Cassel:2009wt, Slatyer:2009vg}, the particles form zero-energy bound states and exhibit resonances. The $\Delta$-dependence is weak.


{\bf \emph{5.\,Signatures \& Constraints.--}} What are the robust signatures of models employing this selection mechanism? The primary signal is a velocity-dependent annihilation rate, but that in itself is not unique to this mechanism. The smoking-gun is that the velocity-dependence is \emph{non-monotonic}: growing as $1/v$ at intermediate $v$, through the competition of $v^2$-suppression of the bare \mbox{$p$-wave} rate and the $\sim1/v^{3}$ Sommerfeld enhancement, and falling off as $v^2$ elsewhere. This means that the constraints from reionization of the CMB should be easily evaded as $v$ is too small, and the signals from galaxy clusters, where $v$ is larger, may be small. The signal may primarily come from intermediate-sized objects such as the Milky Way (MW), nearby galaxies, and dwarf spheroidal galaxies (dSph). Interestingly, these late-time \mbox{$p$-wave} annihilation rates may be significantly larger than the relic annihilation rate.

\begin{figure}[t]
 \includegraphics[width=0.85\columnwidth]{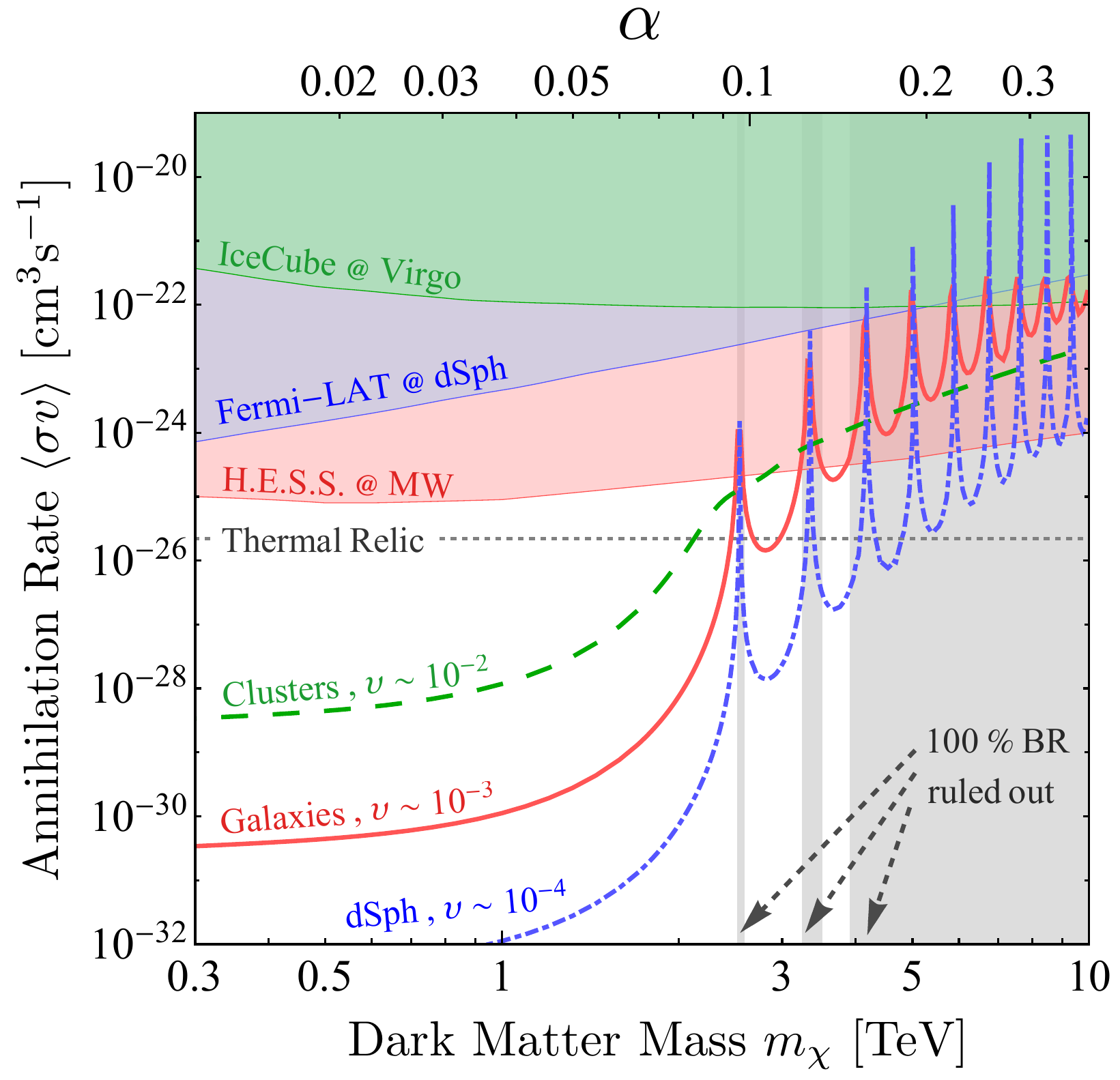}
 \caption{\mbox{$p$-wave} DM annihilation rates in different astrophysical sources and source-specific constraints from indirect detection searches. The annihilation rates have a signature non-monotonic $v$-dependence over and above the resonances, e.g., for $m_\chi>4$\,TeV the galactic annihilation rate (solid line) exceeds that in clusters (dashed line) and dwarf galaxies (dot-dashed line). In DM mass-ranges shown by gray vertical bands a $100\%$ branching ratio to $\mu^+\mu^-$ is ruled out.}
 \label{fig:3}
\end{figure}
Figure\,\ref{fig:3} illustrates the signatures and constraints for this mechanism, through the model described here. We choose a representative mediator mass $m_\rho=30$\,GeV and mass-gap $\Delta=10$\,GeV. The relic density constraint is satisfied everywhere; the dark fine-structure constant $\alpha$ (shown on the upper abscissa) is determined for a given $m_\chi$ (shown on the lower abscissa) by the \mbox{$s$-wave} co-annihilation rate $\pi\alpha^2/(3m_\chi^2)\simeq\langle \sigma v\rangle^{\rm relic}$. At small $m_\chi$ the late-time annihilation rate scales as $v^2$, as expected for \mbox{$p$-wave} annihilations when $S$ is not too large. $\langle \sigma v\rangle$ is the largest in galaxy clusters (dashed line), followed by galaxies (solid line) and dwarf galaxies (dot-dashed line). However, with stronger selective Sommerfeld enhancement, at larger $m_\chi$ the rate is the largest in galaxies. At resonant values of $m_\chi$, it may be the largest in dwarf galaxies. The Sommerfeld factor at recombination is similar to that in dwarfs, both being in the saturated $S$ regime, but the annihilation rate is suppressed by $v^2_{\rm CMB}/v_{\rm dSph}^2\simeq10^{-8}$.

Observation of DM annihilation requires a connection between the dark sector and the visible sector. This is model-dependent and parametrized in the branching ratio $\textrm{BR}$ of the annihilation rate to the specific SM particles. As always, indirect detection constrains $\textrm{BR}\times\langle\sigma v\rangle$. For these models, constraints obtained using one source do not directly apply to another, thanks to the non-monotonic velocity-dependence. Naturally, {velocity-resolved multi-source indirect detection} of the annihilation signal is of key importance here\,\mbox{\cite{Campbell:2011kf, Ng:2013xha, Speckhard:2015eva, Choquette:2016xsw, Powell:2016zbo}}. We compare the predicted rate in each source-class with the limits obtained for that source-class. In Fig.\,\ref{fig:3}, a $100\%$ branching ratio to $\mu^+\mu^-$ is ruled out within the gray vertical bands, due to H.E.S.S. observations of the Milky Way (red shaded region)\,\cite{Abdallah:2016jja}. Constraints from observations of dwarf galaxies, e.g., by Fermi-LAT\,\cite{Ackermann:2015zua}\,(blue shaded region) and AMS-02\,\cite{Ibe:2015tma, Aguilar:2016kjl}, also independently constrain resonant slivers within these bands. Improvements in IceCube observations of the Virgo cluster (green shaded region)\,\mbox{\cite{Dasgupta:2012bd, Murase:2012rd, Aartsen:2013dxa}} and Fermi-LAT observations of the Fornax cluster may be interesting for $m_\chi\simeq$ \mbox{(2-4)\,TeV}\,\cite{Ackermann:2010rg}. CMB  data are significantly less constraining (not shown), than for \mbox{$s$-wave} models. Surprisingly, a purely \mbox{$p$-wave} late-time annihilation rate can be larger than $\langle \sigma v\rangle^{\rm relic}$ and is eminently detectable.

DM has long-range interactions in these models, and constraints on small-scale structure, e.g., from Bullet Cluster, may apply\,\cite{Clowe:2006eq, Buckley:2009in, Peter:2012jh, Viel:2013apy, Schneider:2013wwa, Kim:2016ujt}. For the model parameters considered here, they happen to be weak. Specific models may also be constrained using collider limits on dark-visible mixing\,\cite{Belanger2013340,Giardino2014}. A rather generic prediction of these models is dark radiation $\Delta N_{\rm eff}\gtrsim0.13$\,\cite{Chu:2014lja}, due to the presence of light mediators or their decay into light SM particles, that will be detectable via future CMB observations\,\cite{Abazajian:2013oma, Wu:2014hta}.

{\bf\emph{6.\,Summary \& Outlook.--}} We have pointed out a selection mechanism that leads to large and possibly observable \mbox{$p$-wave} annihilation rates in the present Universe, without enhancing \mbox{$s$-wave} rates. The smoking gun of this mechanism is the signature velocity-dependence and source-dependence of $\langle \sigma v\rangle$, with the possibility of it exceeding $\langle \sigma v\rangle^{\rm relic}$. These features are distinctive of large \mbox{$p$-wave} annihilation of degenerate multi-level DM.

We then discussed a concrete model implementing the selection mechanism and showed that large portions of its parameter space are already probed by existing experiments. The exact constraints are model-dependent, but in general multi-source indirect DM detection, cosmological searches for dark radiation, and small-scale DM structure are the main avenues for testing this mechanism. Collider searches can pin down the dark-to-visible sector connection.

This mechanism opens a new area for model-building and phenomenology, allowing enhanced DM annihilations in specific sources where DM has velocities in an optimal range. As further work, one may also consider the several variations on this theme: more than two DM particles in the dark sector, even-$s$ incoming states, repulsive interactions, multiple mediators, etc. Some of these possibilities may also turn out to be theoretically interesting and find phenomenological application.

{\bf\emph{Acknowledgements.--}} This work was partially funded through a Ramanujan Fellowship of the Dept. of Science and Technology, Government of India, and the Max-Planck-Partnergroup ``Astroparticle Physics'' of the Max-Planck-Gesellschaft awarded to B.D. We acknowledge use of the {\sc FeynCalc} package~\cite{Shtabovenko:2016sxi} and invaluable help from Vladyslav Shtabovenko. We thank Ranjan Laha and Kenny\,C.\,Y.\,Ng for their many useful comments on the manuscript.
\vspace{-0.3cm}
\bibliography{final_draft}
\end{document}